\documentclass[12pt,letterpaper]{article}
\usepackage{amssymb,latexsym,amsmath,amscd,amsthm}
\usepackage{amsfonts}
\usepackage{epsfig}
\usepackage[active]{srcltx}
\usepackage{color}
\usepackage[latin1]{inputenc}
\usepackage[round]{natbib}
\usepackage{xspace}
\usepackage{appendix}
\usepackage{placeins}
\usepackage[a4paper]{geometry}
\usepackage{rotating}
\usepackage{lscape}
\usepackage{caption}

\renewcommand{\section}[1]{%
\bigskip
\begin{center}
\begin{Large}
\normalfont\scshape #1
\medskip
\end{Large}
\end{center}}

\renewcommand{\subsection}[1]{%
\bigskip
\begin{center}
\begin{large}
\normalfont\itshape #1
\end{large}
\end{center}}

\renewcommand{\subsubsection}[1]{%
\vspace{2ex}
\noindent
\textit{#1.}---}

\renewcommand{\tableofcontents}{}

\bibpunct{(}{)}{;}{a}{}{,}  

\newcommand{\eri}{\texttt{Erik+2}\xspace}
\newcommand{\svd}{\texttt{ErikSVD}\xspace}
\newcommand{\nj}{\texttt{NJ}\xspace}
\newcommand{\ml}{\texttt{ML}\xspace}
\newcommand{\mll}{\texttt{ML}}
\renewcommand{\mp}{\texttt{MP}\xspace}
\newcommand{\bppml}{\texttt{bppml}\xspace}

\definecolor{pink}{rgb}{1,0,1}

\newif\ifprivate
\privatefalse

\def\???{\ifprivate {\bf {???}} \marginpar{{\Huge {\bf ?}}}
\else \fi}

\begin{document}

\bigskip

\bigskip
\medskip
\begin{center}

\noindent{\Large \bf Invariant versus classical quartet inference when evolution is heterogeneous across sites and lineages}
\bigskip



\noindent {\normalsize \sc Jes\'{u}s Fern\'{a}ndez-S\'{a}nchez$^1$ and Marta Casanellas$^1$}\\
\noindent {\small \it
$^1$Dept. Matem\`{a}tica Aplicada I, Universitat Polit\`{e}cnica de Catalunya}\\
\end{center}
\medskip
\noindent Dept. Matem\`{a}tica Aplicada I, 
Universitat Polit\`{e}cnica de Catalunya \\ Gran Via 585, 08023-Barcelona, Spain \\ 
E-mail: jesus.fernandez.sanchez@upc.edu, marta.casanellas@upc.edu.\\



\subsubsection{Abstract}
One reason why classical phylogenetic reconstruction methods fail to correctly infer the underlying topology is because they assume oversimplified models. In this paper we propose a topology reconstruction method consistent with the most general Markov model of nucleotide substitution, which can also deal with data coming from mixtures on the same topology. It is based on an idea of Eriksson on using phylogenetic invariants and provides a system of weights that can be used as input of quartet-based methods.
We study its performance on real data and on a wide range of simulated 4-taxon data (both time-homogeneous and nonhomogeneous,  with or without among-site rate heterogeneity, and with  different branch length settings). We compare it to the classical methods of neighbor-joining (with paralinear distance), maximum likelihood (with different underlying models), and maximum parsimony. Our results show that this method is accurate and robust, has a similar performance to ML when data satisfies the assumptions of both methods,  and outperforms all methods when these are based on inappropriate substitution models or when both long and short branches are present. If alignments are long enough, then it also outperforms other methods when some of its assumptions are violated.\\
\noindent[\textbf{Keywords}:  phylogenetic invariants, topology reconstruction, general Markov model, heterogeneity across lineages, heterogeneity across sites, yeast]\\




\section{Introduction}
Classical methods of phylogenetic tree topology reconstruction  are known to have limitations. For example, maximum likelihood (\ml) is known to fail when data violates some of the underlying model assumptions \citep{Swofford2001, Kuck2012, Ho2004}; maximum parsimony (\mp) is statistically inconsistent in the Felsenstein zone \citep{felsenstein1978}; and neighbor-joining (\nj) is subject to the choice of an unbiased distance and it is not as accurate as \ml when both methods can be applied \citep{Tateno1994}. When trying to estimate distant phylogenies, neglecting 
heterogeneity in the substitution process across lineages (HAL from now on, as denoted in \citet{jayaswal2014}) 
or heterogeneity  across sites (HAS) may result in inaccurate phylogenetic estimates
(see \citet{Yang1995,Ho2004,foster2004, galtier1998, felsenstein1978, yang1994, fitch1986, Stefankovic2007, Kolaczkowski2004} among others).   


%
\textit{Phylogenetic invariants} were first introduced by  \citet{cavender1987} and  \citet{lake1987} as a non-parametric method of phylogenetic reconstruction: they are equations satisfied by any possible joint distribution of character patterns at the leaves of a tree evolving under an evolutionary Markov model. The potential of phylogenetic invariants was the ability of dealing with more general models  and of detecting the topology without estimating branch lengths or substitution parameters \citep[see][chapter 22]{Felsenstein2004}. In particular, they can handle HAL
better than other methods \citep{casanellas2007,holland2013} and (some) could deal with
HAS, as  Lake's invariants did \citep{lake1987}. Nevertheless, only a few phylogenetic invariants
were known by that time, it was not clear how to use them  \citep{Felsenstein2004}, they seemed useless for large trees, and the approach was laid aside by the bad results obtained in simulations \citep{huelsenbeck1995}.

\citet{eriksson2005} proposed a new topology reconstruction method, \svd, based on the work on invariants of
\citet{AllmanRhodeschapter4}. The underlying idea is  that organizing the joint distribution of character patterns
according to a bipartition of the set of taxa gives a \textit{bipartition matrix} (\ref{flattening}) of rank $\leq 4$ if the bipartition is induced by an edge of the tree (and otherwise, the rank is higher). This result holds for any set of  DNA
sequences evolving under the most general Markov model (GMM), also known as Barry Hartigan's model \citep{barryhartigan87trans, allman2008, jayaswal2005}. This is the most general HAL model as
it allows different instantaneous
rate matrices and heterogeneous composition at different parts of the tree, even locally along each branch \citep{Jayaswal2011}.
\svd does not use  phylogenetic invariants directly but computes the Frobenius distance of the bipartition matrices to the set of matrices of rank $\leq 4$.
Although it is nowadays clear that phylogenetic invariants derived from rank conditions on matrices
are the only relevant invariants for reconstructing  the topology \citep{casfer2010},
the original method \svd turned out not to be
accurate enough to compete against standard methods \citep{eriksson2005}, especially in the presence of long branches and short alignments.

Here we revisit \svd by correcting the target matrix: in the method \eri proposed here we consider the two possible transition matrices from the
states of one side of the bipartition to the other (that is, we normalize by column and row sum the
bipartition matrix of \svd). This correction is made to take into account that the rank of the bipartition matrix obtained from empirical distributions  could be affected by the presence of  long-branch attraction situations
(see Appendix 1). The original
\svd was already \textit{statistically consistent} (that is, as the empirical distribution approaches the theoretical distribution, the probability of correctly reconstructing the tree goes to one) and  so is the new method (see Materials and Methods). 

%
%

\eri is model-based as it assumes a general Markov model of evolution (and it could also be redesigned to incorporate more restrictive Markov models or even aminoacid substitution models),
but is non-parametric in the sense that it does not attempt to recover the parameters of the model.  Moreover, the theoretical background of \eri permits to apply it on HAS data evolved on the same tree topology under GMM \citep{jayaswal2014}: that is, a parameter $m$ can be introduced so that \eri considers the sites of the alignment to be divided into  $m$ categories, each  evolving on the same topology but with (possibly) different Markov substitution matrices --this is called an \textit{m-mixture}  \citep{Stefankovic2007}. For example, discrete-gamma rates or the heterogeneous tree in \citep{Kolaczkowski2004} are instances of mixtures, and \ml is known to fail under these conditions even when consistent underlying homogeneous models are considered \citep{Kuck2012,Kolaczkowski2004}. For $m$-mixtures, the rank of the bipartition matrix induced by an edge is not larger than $4m$ \citep[e.g.][]{rhodessullivant} so that in this case we use the distance to matrices of rank $\leq 4m$.

We develop \eri on 4-taxon trees and study its performance on simulated and real data. Using computer simulations we compare it to the classical methods \ml, \nj, \mp and to the original \svd in many different scenarios. We chose quartets because
they are the smallest building blocks of phylogenetic reconstruction \citep{Ranwez2001} and they
are widely used as a hint of efficiency and robustness of the method under study \citep{huelsenbeck1995}. \eri evaluates the three possible quartet topologies and returns a system of weights that can be used as input for quartet-based methods (see the Methods section).

Some of our computer simulations are generated under the general Markov process that underlies \eri
and some are based on themost general time-reversible (GTR) and homogeneous across lineages model (homGTR from now on).  We also simulate HAS data by generating either 2-mixtures on the same topology evolving under GMM or Gamma continuously-distributed rates across sites under the homGTR model. 
Throughout the paper \nj has been considered with the paralinear distance,
and \ml computations have been based on continuous-time models (with parameters to be estimated by the method)
considering homogeneity or heterogeneity across lineages and sites depending on the situation (we detail it explicitly in figure captions).

The performance of \eri on real data is analyzed on the eight species of yeast studied in \citet{Rokas2003} with the concatenated alignment provided by \citet{jayaswal2014}. We investigate whether the quartets output by \eri and \svd support the tree $T$ of \citet{Rokas2003} or the alternative tree $T'$ of \citet{Phillips2004}, and the mixture model proposed by \citet{jayaswal2014}.

\FloatBarrier
\section{Results}
We present the performance of the new method \eri, the original method of Eriksson (\svd), and the classical
methods \ml, \nj and \mp, on quartet reconstruction on different simulated data. 
\eri is publicly available at the webpage \newline
http://www.pagines.ma1.upc.edu/$\sim$casanellas/Erik+2.html.

\subsection{Homogeneity across sites}
\begin{figure}[h]
\begin{center}
 \includegraphics[scale=0.6]{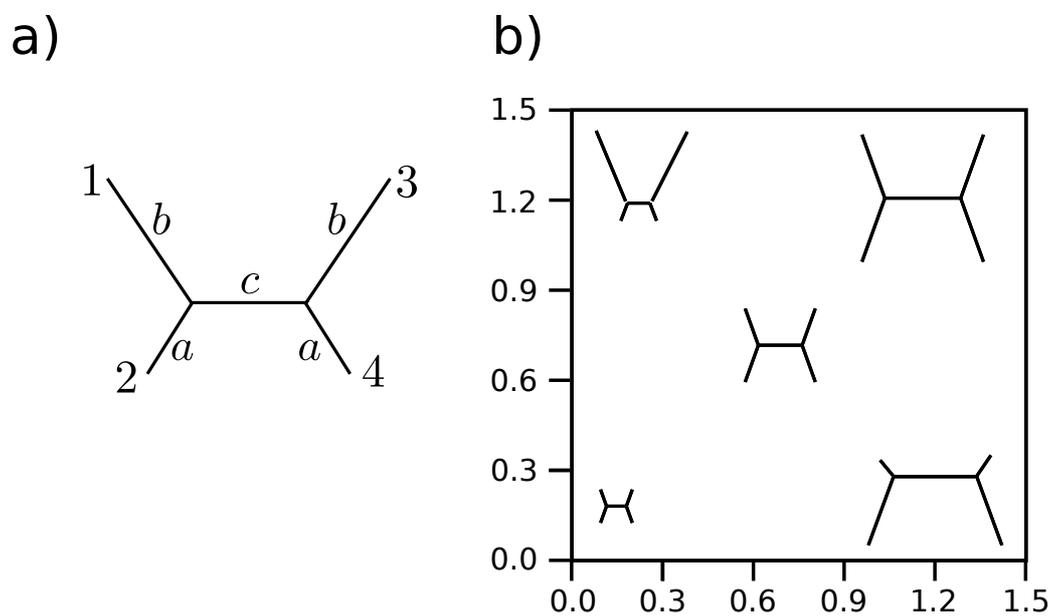}
\end{center}
\linespread{1.3}
\footnotesize
\caption{\label{tree} 
a) 4-leaf tree where the length of two opposite branches is
represented by $a$; the other two peripheral branches have length $b$; and the length of the interior
branch is denoted by $c$. Branch lengths will be measured in the expected number of substitutions per site. b) Tree space used in Figure 3 and 4: on the left tree,  branch length $c$ is set equal to $a$ and branch lengths $a$ and $b$ are varied from 0.01 to 1.5 in steps of 0.02.}
\end{figure}

\begin{figure}[h]
\begin{center}
 \includegraphics[scale=0.7]{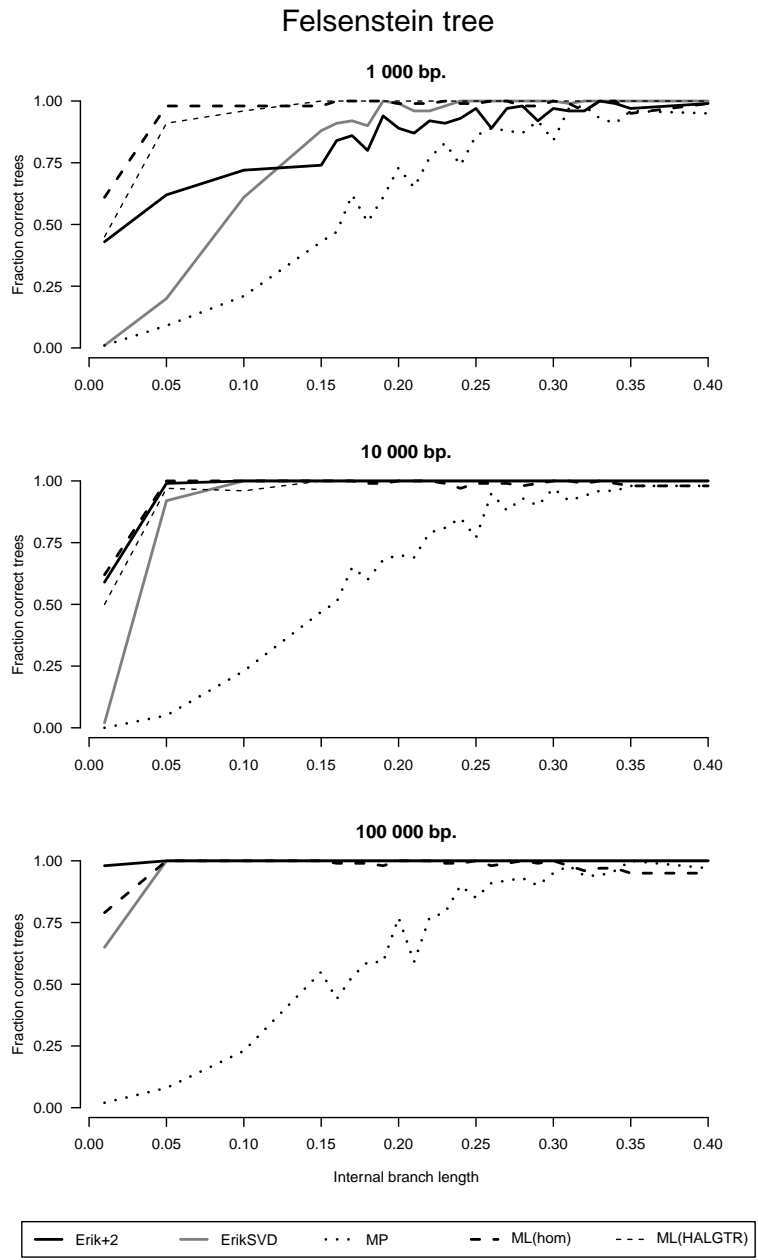}
\end{center}
\linespread{1.3}
\footnotesize
\caption{\label{single} 
Percentage of correctly reconstructed topologies by \eri, \svd, maximum likelihood \ml, and maximum-parsimony \mp on data
generated under the general Markov model (GMM) on the tree of Figure 1.a with $a= 0.05$, $b=0.75$, and varying the internal
branch length $c$. Two types of  \ml inference have been applied here: \mll(hom) estimating the most general homogeneous continuous-time model, and \mll(HALGTR) estimating a HAL GTR model (due to the time of execution of this last method, we could only test it for 1~000 bp. and 10~000 bp.).
}
\end{figure}

%
First of all we consider a tree subject to long branch attraction. On the tree of Figure \ref{tree}.a 
we fix $a= 0.05$, $b=0.75,$
and let the internal branch length $c$ vary in the range $[0.01,0.4]$ so that the tree lies in the Felsenstein zone.
Alignments of lengths 1~000, 10~000 and 100~000 base pairs (\emph{bp}.) were generated under GMM according to this tree.
The results obtained for \eri, \ml , \mp and \svd on these data are shown in Figure \ref{single}. In this figure 
two models underlying \ml computations have been considered: 
the most general homogeneous continuous-time model, \mll(hom) from now on, and a HAL GTR model, \mll(HALGTR) henceforth. \ml has a similar performance with both models.
We observe that, \eri is more accurate than \svd in general and
especially when the
interior branch length is short 
(only for length 1~000 and $c\in(0.13,0.25)$ \svd outperforms slightly \eri). For 1~000 sites, both versions of \ml perform better than \eri, but when more data is available 
\eri outperforms \ml (in its both versions). Notice incidentally that the accuracy of \ml or \mp does not seem to increase as the length of the alignment grows, while it certainly does for \svd and \eri. 

\begin{figure}[h]
\begin{center}
 \includegraphics[scale=0.5]{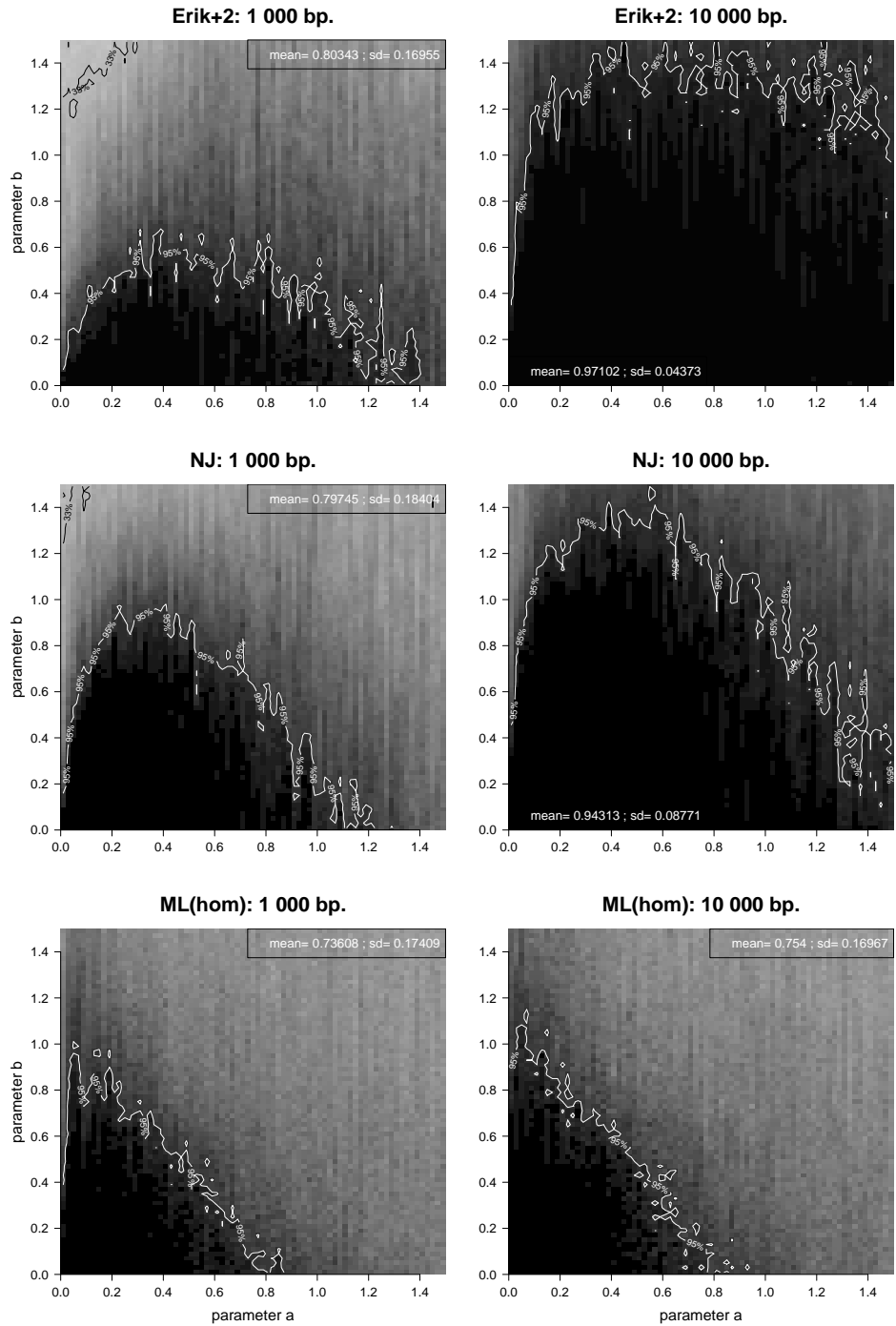}
\end{center}
\linespread{1.3}
\footnotesize
\caption{\label{treespaceGMM} 
Performance on the tree space of Figure 1.b on data generated under the
general Markov model: black is used to  represent 100 \% of successful topology reconstruction, white to represent 0 \%, and different tones of gray the intermediate frequencies.  \emph{Top}: \eri; \emph{Middle}: Neighbor-Joining (paralinear distance); \emph{Bottom}: \mll(hom) estimating
the most general homogeneous across lineages continuous-time model. \emph{Left}: 1 000 bp; \emph{Right}: 10 000 bp.
}
\end{figure}

\begin{figure}[h]
\begin{center}
 \includegraphics[scale=0.5]{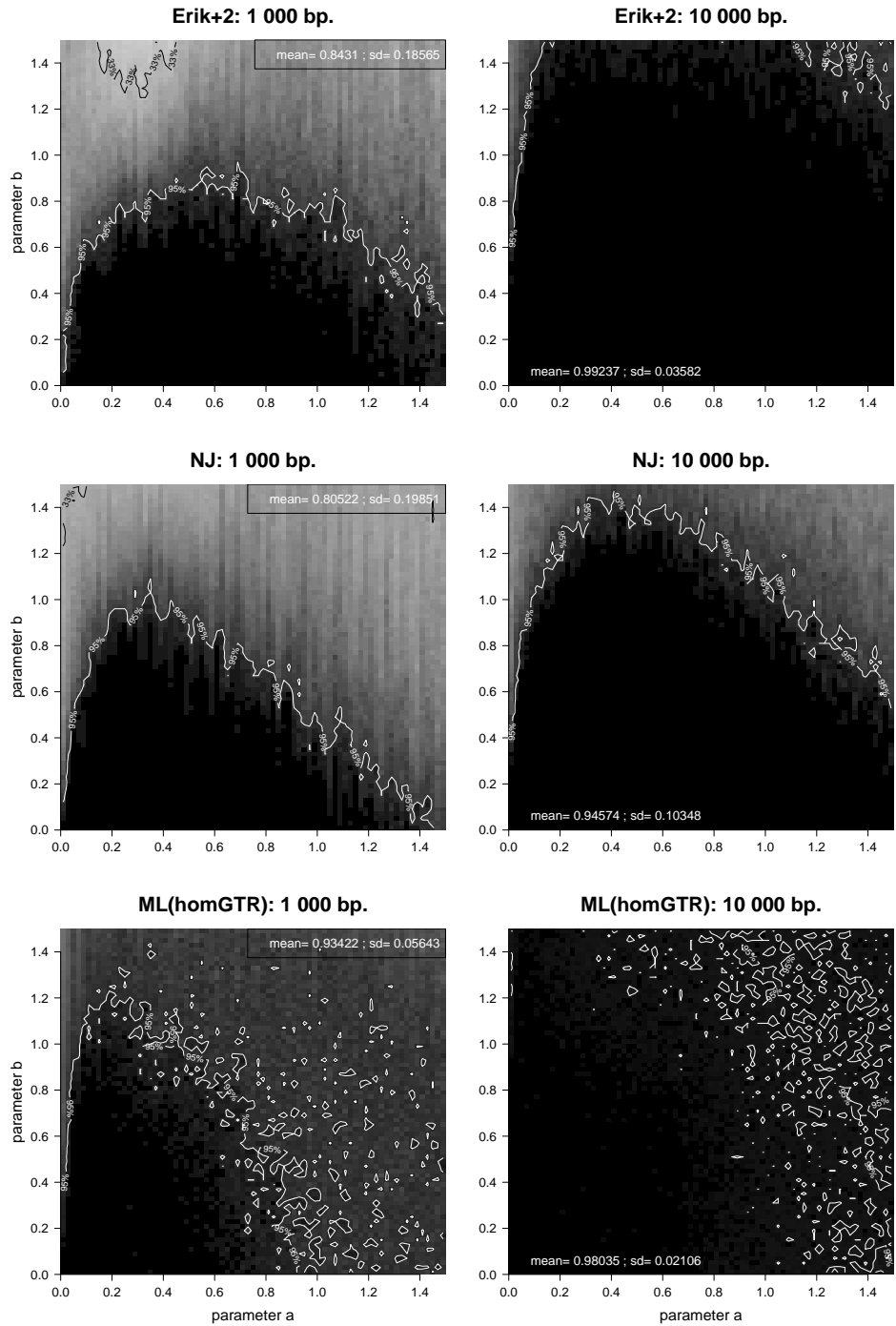}
\end{center}
\linespread{1.3}
\footnotesize
\caption{\label{treespaceGTR}
Performance on the tree space of Figure 1.b  on data generated under the (homogeneous across lineages) GTR
model: black is used to  represent 100 \% of successful topology reconstruction, white to represent 0 \%, and different tones of gray the intermediate frequencies. \emph{Top}: \eri; \emph{Middle}: Neighbor-Joining (paralinear distance); \emph{Bottom}: \mll(homGTR) estimating
homogeneous GTR model. \emph{Left}: 1 000 bp; \emph{Right}: 10 000 bp.}
\end{figure}

In a more complete study, we adopted a similar approach to \citet{huelsenbeck1995} to test
different methods. More precisely, we evaluate the methods on a \textit{tree space} (see Figure \ref{tree}.b) where the quartets are as in Figure \ref{tree}.a with $c=a$, and the branch lengths $a$ and $b$ vary between 0 and 1.5 in steps of 0.02.
For each pair $a,b$ we generated 100 alignments of a fixed length and represented in a gray scale the success of different methods in recovering the right topology (black means 100 \% of success, and white 0 \%).
%
The methods \eri, \ml, and \nj with paralinear distance have been tested according to this approach.
The results for 1~000 and 10~000 bp. are shown in Figure \ref{treespaceGMM}  for data generated under GMM and \ml estimating the most general homogeneous continuous-time model \mll(hom), and in Figure \ref{treespaceGTR} for data generated under homogeneous GTR model and \ml estimating exactly the same model, \mll(homGTR).

In Figure \ref{treespaceGMM}, we see that both \mll(hom) and
\nj have lower accuracy than \eri, as it was expected under data that violates the assumptions of \ml and \nj.
In both figures \ref{treespaceGMM} and \ref{treespaceGTR} we observe that, while \eri and \nj drastically increase their accuracy when alignment length is multiplied by
10, \ml does not significantly improve with alignment length (especially in Figure \ref{treespaceGMM} when the substitution model assumed by \ml is incorrect).
In Figure S1 of the Appendix 2 the reader can find  the performance of \svd on the tree space of Figure \ref{tree}.b, confirming the improvement of \eri over the original method. The average success and standard deviation achieved by these methods on this tree space are shown in Table \ref{tab:mean_sd} (where alignment length 500 bp. is also included).
%

%

%


It is worth pointing out that, for alignments of 1~000 bp. evolving under homogeneous GTR model, \mll(homGTR) seems to perform better than \eri in the Felsenstein zone. However, for length 10~000, \eri already outperforms \mll(homGTR) (Fig. \ref{treespaceGTR}). Moreover, the global accuracy of \mll(hom) drastically drops when applied to data obtained under GMM (see Fig. \ref{treespaceGMM}).
Notice also that whereas the accuracy of \nj and \ml
drops when all branches are long (top right corner), the performance of \eri seems less sensitive to long branches.
%
%

We have also evaluated the version of \eri with 2-mixtures ($m=2$) on the same data (see Fig. S3 in Appendix 2).  The accuracy obtained for alignments of 1~000 bp. is similar to that of \eri with $m=1$ (the means are 0.790 and 0.803, respectively), and hence the choice $m=2$ appears as a good option when alignments are long enough and we ignore whether the data comes from mixtures or not (see also Fig.
S4 in Appendix~2).

\subsection{Heterogeneity across sites (HAS)}
\begin{figure}
\begin{center}
\includegraphics[scale=0.7]{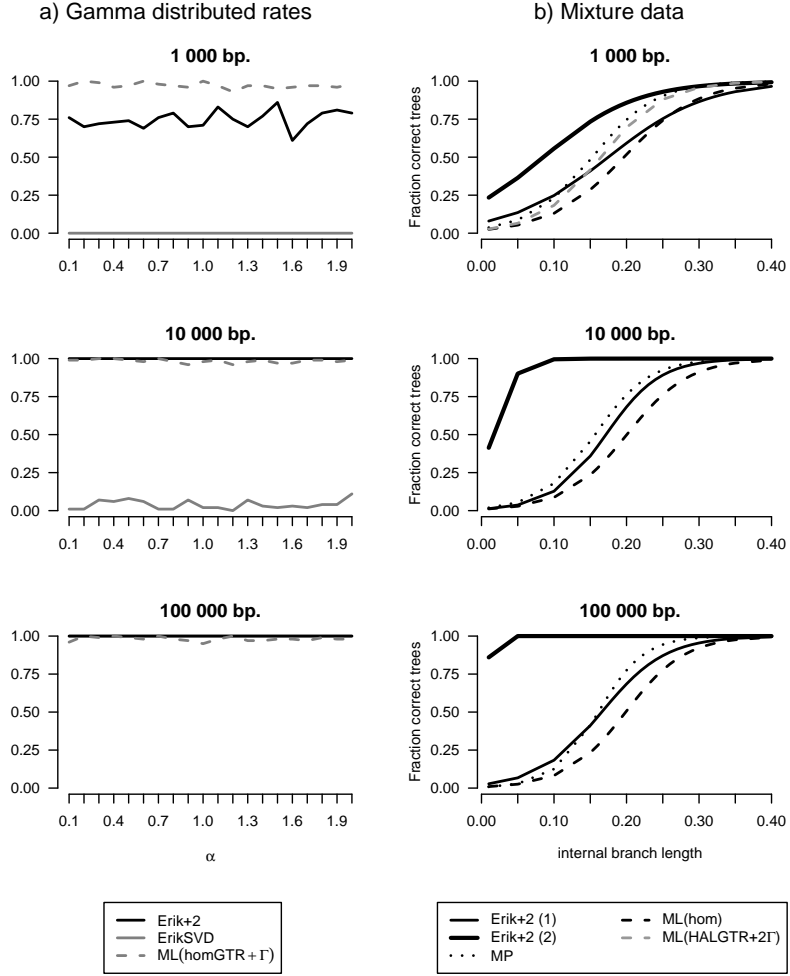}
\end{center}
\vspace{-1cm}
\linespread{1.3}
\footnotesize
\caption{\label{2partitions} 
Percentage of correctly reconstructed topologies by  different methods on alignments of
lengths 1 000, 10 000 and 100 000 bp. shown from top to bottom.  
a) \eri, \svd, and \ml estimating the GTR model (homogeneous across lineages) with auto-discrete Gamma-rates and denoted as \mll(GTR+$\Gamma$).  Data simulated under (homogeneous across lineages) GTR model with continuous Gamma-rates and parameter $\alpha$ varying between 0.1 and 2 on the tree of Figure 1.a with branch lengths $a=c=0.05$ and $b=0.75$ (\mp had 0\% success on this data, so we do not show it).
b) Data generated under GMM with 2
categories according to the test designed in (Kolaczkowski and Thornton 2004), varying the internal branch length, and recovering with \eri with $m=2$, \eri with $m=1$, \mp, \mll(hom) estimating the most general homogeneous across lineages model, and \mll(HALGTR+2$\Gamma$) estimating HAL time-reversible model with 2 discrete-Gamma categories (due to the time of execution of this last method, we could only test it for 1~000 bp). The plot represents the logistic regression curve of the output of each method. In all cases, \ml had to estimate all parameters. 
}
\end{figure}

%
On the same tree of Figure \ref{tree}.a, 
we
generated data under homogeneous GTR model with sites varying according to a Gamma distribution with parameter $\alpha=\beta$ in the range $(0,2]$ varying in steps of $0.1$. Small values of this parameter indicate a lot of variation
across sites \citep{yang1993}. While this setting  violates the hypotheses of the model underlying \eri and \svd, in this case  maximum likelihood is estimating a homogeneous GTR model with rates varying according to the auto-discrete Gamma model  \mll(homGTR+$\Gamma$) \citep{yang1994}.
The results appear in Figure \ref{2partitions}.a, where we observe that \eri manages to overcome the violation of its hypotheses giving 100\% success
already for 10~000 bp., while \svd gives notably worse results. \ml is more successful than \eri for 1~000 bp., but both methods have a similar performance on longer alignments.
On the same data we also tested \mp, obtaining in all cases the incorrect tree $13|24$ (and therefore we do not represent the corresponding 0\% line in the figure).

As mentioned above, one of the main features of \eri is that it can deal with different categories of 
evolutionary rates. In order to test its accuracy on such setting,  we used the approach  of \citet{Kolaczkowski2004}. We considered two categories of the
same size both evolving under GMM on the tree of Figure \ref{tree}.a: the first category corresponds to branch lengths $a=0.05$, $b=0.75$, while the second corresponds to $a=0.75$ and $b=0.05$. The internal branch length was set to the
same value in both categories and varied from 0.01 to 0.4. 
In Figure \ref{2partitions}.b) we present the performance of \eri (with $m=1$ and $m=2$), \mp, \mll(hom), and \ml estimating a HAL GTR model with discrete Gamma rates with 2 categories, \mll(HALGTR+2$\Gamma$) henceforth.
We included \mp in this study because, as stated in \cite{Kolaczkowski2004}, it performs better than \ml estimating a single category model. This claim is confirmed by the results in our simulations with both versions of \ml.
It is worth pointing out that even \eri with $m=1$ performs better than \mll(hom) for internal branch length $\leq 0.25$, and than \mll(HALGTR+2$\Gamma$)  for internal branch length $\leq 0.15$.  Also, notice
that for length 10~000 and larger, the accuracy of \eri with $m=2$ is always greater than 33\%, even if the internal branch length is close to zero. This does not happen for \ml, \mp, which are clearly inconsistent in this setting.


\subsection{Performance on real data}
We considered the data provided by \citet{jayaswal2014} with 42~337 second codon positions of 106 orthologous genes of \emph{Saccharomyces
cerevisiae}, \emph{S. paradoxus}, \emph{S. mikatae}, \emph{S. kudriavzevii}, \emph{S. castellii}, \emph{S. kluyveri}, \emph{S. bayanus}, and \emph{Candida albicans}. The phylogenetic tree of these species was originally studied in \citet{Rokas2003}, where a tree topology $T$ was identified with 100\% bootstrap support for the concatenated alignment of these genes. This tree is widely accepted by the community but its correct inference is known to depend on  the consideration of HAL \citep{Rokas2003, Phillips2004, jayaswal2014}. For example, \citet{Phillips2004} obtain an alternative tree $T'$ with 100\% bootstrap using the method of minimum evolution, but identified the incorrect handling of compositional bias as responsible for this inconsistency. Moreover, according to \citet{jayaswal2014} these data is best modeled by taking into account HAL plus two different rate categories and invariable sites. In our setting, this would involve three mixtures. We apply \svd and \eri with $m=1,2,3$ to 4-taxon subalignments and investigate the proportion of output quartets that are compatible with $T$ or $T'$. The results displayed in Table \ref{tab:realdata} show that \eri supports the tree $T$ and the model suggested by \citet{jayaswal2014} ($m=3$), whereas \svd gives more support to the alternative tree $T'$.

\subsection{Time of execution}
We have compared the time of execution of the different reconstruction methods used in our simulations with 100 alignments of length 1~000 bp on a 3.2GHz processor.
The results obtained show that \nj is the fastest method, 1.324s. \svd and \eri take 1.928s and 2.148s respectively, and \mp takes 3.984s. Finally, \ml is the slowest method by far because it has to infer the model parameters: using PAML software, \mll(hom) and \mll(homGTR)  need about 10 seconds, and using \bppml of Bio++ package, \mll(HALGTR) and \mll(HALGTR+2$\Gamma$) need about 200 minutes.   

\section{Discussion}
The simulation studies  show that \eri is an accurate and robust topology reconstruction method on quartets, especially in situations where other methods systematically fail (compositional heterogeneity and/or rate heterogeneity across lineages, or long branch attraction). In such scenarios, \eri outperforms the method of Eriksson, \svd, and classical methods like \mp, \nj and \ml based on models that cannot accommodate these assumptions. \eri is based on the most general Markov model and hence accounts for HAL data, even locally at each edge. When its assumptions are violated, for example in the presence of continuous Gamma-distributed rates among sites, we have shown that it is highly accurate if there is enough data. As observed, \eri can also deal with $m$-mixtures on the same tree topology (although for quartets the limit is $m=3$).  Even more, using \eri with  $m=2$ is probably the best option for large alignments 
when mixed/unmixed nature of data is unknown.

On the experiments we presented, the overall performance of \ml is quite accurate if model assumptions are not violated, confirming the conclusions of \citet{Kolaczkowski2004, Kuck2012}. Also in line with these papers, we corroborate that long sequences do not improve \ml performance on data that do not satisfy the hypothesis of the underlying model. Moreover, \ml is by far the slowest among the methods tested here,
while  \eri  is slightly twice slower than \nj. Another drawback of \ml is that, quite often, it does not converge when it is computed on the incorrect topology, which makes the comparison of likelihoods impossible. Whereas the goal of \eri is to reconstruct the topology, \ml is designed to estimate the parameters of the substitution matrices and,  it would probably be a good choice to use first \eri and then \ml to estimate the parameters.
In our simulation study, \nj (with paralinear distance) and \mp have been the methods with least success, which is not so surprising if one takes into account that they are also the less adaptable  to general data.


We have only developed \eri for quartets with the aim of validating it as a successful method, and it is still a work in progress to further develop it for larger number of taxa. Using \eri to evaluate the confidence of particular bipartitions of large sets of taxa is already a viable option, and in this case one can deal with a larger number $m$ of categories (the maximum $m$ allowed depends on the size of the subsets $A$, $B$ of taxa involved in the bipartition: $4^{min\{|A|,|B|\}-1}-1$). We have also started testing its weights as input of weighted quartet-based methods with high success (unpublished). In particular, it outperforms global \nj, which makes \eri a potential input method for quartet-based methods  \citep{stjohn2003}.


\newpage

\FloatBarrier
\section{Materials and Methods}

\subsection{\svd and \eri methods}
\eri arises as a variation of the method described by \citet{eriksson2005} by normalizing certain bipartition matrices obtained from an alignment of nucleotide sequences. As in the original method, the information contained in the alignment is recorded as a vector $\tilde{p}$ whose coordinates are the observed relative frequencies of possible patterns at the leaves. In the case of an alignment of four taxa 1,2,3,4, each possible (trivalent) topology is determined by a bipartition of thetaxa: $12|34$, $13|24$ or $14|23$. For each bipartition $A|B$, a matrix $M_{A\mid B}(\tilde{p})$ is considered by rearranging the coordinates of $\tilde{p}$ according to it, so that the rows of the matrix $M_{12|34}(\tilde{p})$ are indexed by all possible observations $(x_1,x_2)$ at the leaves $1,2$, and similarly for columns and observations $(x_3,x_4)$ at the leaves $3,4$.
For example, the $(AG,CT)$-entry of $M_{12|34}(\tilde{p})$ is the relative frequency $\tilde{p}_{AGCT}$ of the pattern $AGCT$ in the alignment. The same entry in $M_{13|24}(\tilde{p})$ corresponds to the relative frequency $\tilde{p}_{ACGT}$ of $ACGT$.
\begin{eqnarray}\label{flattening}
 M_{12\mid 34}(\tilde{p})=\left (
 \begin{array}{ccccc}
 \tilde{p}_{AAAA} & \tilde{p}_{AAAC} & \tilde{p}_{AAAG} & \ldots & \tilde{p}_{AATT} \\
 \tilde{p}_{ACAA} & \tilde{p}_{ACAC} & \tilde{p}_{ACAG} & \ldots & \tilde{p}_{ACTT}\\
 \tilde{p}_{AGAA} & \tilde{p}_{AGAC} & \tilde{p}_{AGAG} & \ldots & \tilde{p}_{AGTT}\\
 \ldots  & \ldots  & \ldots  &  & \ldots \\
  \end{array}
 \right )
\end{eqnarray}

Assume that the coordinates of $\tilde{p}$ are the empirical estimates of the theoretical joint distribution $p$ at the leaves of a tree $T$ evolving under GMM, say $T=12|34$. Then 
the key point is Theorem 19.5 of \citet{eriksson2005} (see \citet{casfer2010} for a complete proof) that claims that 
the rank of $M_{A|B}(p)$ is $4$ if $A|B=12|34$, and  $4^2$ otherwise (if the substitution matrices that generated $p$ were general enough). Eriksson's idea is to
compute the \emph{Frobenius distance} \citep[that is, the euclidean distance if we view the matrices as elements in $\mathbb{R}^{4^2\times 4^2}$,][]{Demmel} $\mathrm{d_4}$  of
the three matrices $M_{12|34}(\tilde{p})$, $M_{13|24}(\tilde{p})$ and $M_{14|23}(\tilde{p})$ to the space of matrices of rank $\leq 4$. In this manner, one derives which of the three matrices is closer to having rank $\leq 4$. The Frobenius distance of a matrix $A$ to $k$-rank matrices, $\mathrm{d_k}(A)$, is easily computed in terms of the singular values of $A$  \citep{Eckart1936}.
%

The main motivation for the variation introduced in \eri arises from the observation that the presence of short branches may seriously affect this distance when its computed from short alignments.
For example, taking the tree of Figure 1.a with small $a$ and large $b$ (a tree corresponding to the so-called Felsenstein's zone), the distance of $M_{13|24}(\tilde{p})$ to 4-rank matrices is smaller than that of $M_{12|34}(\tilde{p})$ (see Appendix 1 for an example). The reason is that a small $a$ implies that the probability of observing the same nucleotide at leaves 2 and 4 is high,
so columns in $M_{13|24}(\tilde{p})$ indexed by $AA$, $CC$, $GG$, or $TT$ capture most of the non-zero entries in the matrix, while other columns may only have few nonzero entries. This makes the matrix to be close to a rank $\leq 4$ matrix, even if $13|24$ is not the correct topology.
By dividing any non-zero column by the sum of its entries, we make all the non-zero columns to have the same weight. As the same situation may occur with rows, we also need to correct the matrix by row sums. 
In this way, each matrix $M_{A|B}(\tilde{p})$ gives rise to a pair of \emph{transition} matrices $M_{A\rightarrow B}(\tilde{p})$ and $M_{A\leftarrow B}(\tilde{p})$, obtained by column and row sum correction, respectively:

\begin{footnotesize}
\begin{eqnarray*}\label{transmat}
 M_{12\rightarrow 34}(\tilde{p})=\left (
 \begin{array}{cccc}
 \frac{\tilde{p}_{AAAA}}{\tilde{p}_{AA++}} &  \frac{\tilde{p}_{AAAC}}{\tilde{p}_{AA++}} &   \ldots &  \frac{\tilde{p}_{AATT}}{\tilde{p}_{AA++}} \\
  \frac{\tilde{p}_{ACAA}}{\tilde{p}_{AC++}} &  \frac{\tilde{p}_{ACAC}}{\tilde{p}_{AC++}} &   \ldots &  \frac{\tilde{p}_{ACTT}}{\tilde{p}_{AC++}}\\
  \frac{\tilde{p}_{AGAA}}{\tilde{p}_{AG++}} &  \frac{\tilde{p}_{AGAC}}{\tilde{p}_{AG++}} &   \ldots &  \frac{\tilde{p}_{AGTT}}{\tilde{p}_{AG++}}\\
 \ldots  & \ldots  &  & \ldots \\
  \end{array}
 \right ) \quad
 M_{12\leftarrow 34}(\tilde{p})=\left (
 \begin{array}{cccc}
 \frac{\tilde{p}_{AAAA}}{\tilde{p}_{++AA}} &  \frac{\tilde{p}_{AAAC}}{\tilde{p}_{++AC}} &   \ldots &  \frac{\tilde{p}_{AATT}}{\tilde{p}_{++TT}} \\
  \frac{\tilde{p}_{ACAA}}{\tilde{p}_{++AA}} &  \frac{\tilde{p}_{ACAC}}{\tilde{p}_{++AC}} &   \ldots &  \frac{\tilde{p}_{ACTT}}{\tilde{p}_{++TT}}\\
  \frac{\tilde{p}_{AGAA}}{\tilde{p}_{++AA}} &  \frac{\tilde{p}_{AGAC}}{\tilde{p}_{++AC}} &   \ldots &  \frac{\tilde{p}_{AGTT}}{\tilde{p}_{++TT}}\\
 \ldots  & \ldots  &  & \ldots \\
  \end{array}
 \right ).
\end{eqnarray*}
\end{footnotesize}
We give a score to any tree $T_{A|B}$ as
\begin{eqnarray*}\label{formula_cft}
\mathrm{sc}(T_{A\mid B}):=\frac{\mathrm{d_4}\left(M_{A\rightarrow B}(\hat{p}) \right) + \mathrm{d_4}\left(M_{B\rightarrow A}(\hat{p}) \right)}{2}.
 \end{eqnarray*}
Notice that  the smaller the score is, the more reliable the topology $T_{A|B}$ is and \eri outputs as correct tree the topology with smallest score.
As the empirical distribution $\hat{p}$ approaches the theoretical distribution $p$, the transition matrices $M_{A\rightarrow B}(\hat{p})$ and $M_{B\rightarrow A}(\hat{p})$  approach the theoretical transition matrices. These have rank 4 for the correct topology because they have the same rank as the theoretical bipartition matrices (as they are obtained from them by dividing rows/columns by scalars). Therefore $\mathrm{d_4}\left(M_{A\rightarrow B}(\hat{p}) \right)$  and $\mathrm{d_4}\left(M_{B\rightarrow A}(\hat{p})\right)$ tend to 0 when $\hat{p}$ approaches the theoretical distribution (as the Frobenius distance is a continuous function) and thus \eri  is statistically consistent.

\eri provides also normalized weights that can be used into weighted quartet-based methods.
Indeed, the score above is turned into a confidence weight by inverting it and normalizing so that the overall sum of weights is 1:
 \[w(T_{A|B}):=\frac{\mathrm{sc}(T_{A|B})^{-1}}{\sum_{T\in \mathcal{T}_4}\mathrm{sc}(T)^{-1}}.\]
%
%
%
%
%
%
The basic model underlying \eri and \svd assumes that all sites in the alignment evolve independently and identically distributed according to a general Markov model. There is no extra assumption  about the shape of substitution matrices (nor stationarity, nor time-reversibility, nor global or local homogeneity). But in \eri we relax the i.i.d hypotheses and allow HAS by considering mixtures in the sense of \citep{Kolaczkowski2004} and  \citep{Stefankovic2007}. That is, a single tree topology $T$ is considered but we allow $m$ categories of Markov processes on $T$ defined by $m$ sets $(\sigma_1,\dots,\sigma_m)$ of substitution parameters.
The proportion of sites contributed by the
$i$-th tree $(T,\sigma_i)$ is denoted by $p_i$ and the joint distribution at the leaves of $T$  follows an $m$-\textit{mixture distribution}: $\sum_i p_i P(T,\sigma_i)$. 
A parameter $m \in \{1,2,3\}$ can be passed to \eri to adapt the method to consider $m$ categories (in this case, we compute the distance $d_{4m}$ to matrices of rank $\leq 4m$). The restriction to 3 categories at most is only due to theoretical results about non-identifiability for quartet trees with four or more partitions (there would be 255 parameters in a 4-mixture, which already fills the whole space of pattern distributions, see \citet{casferked}).


We had also developed different modifications of the original method of Eriksson, all of them
showing lower success than the version considered here. Therefore in this paper we only present the
results corresponding to \eri. 

\subsection{\ml and \nj}
Software PAML \citep{yang1997} was used to estimate the likelihood under time-homogeneous models. Depending on the simulations we used either the most general continuous-time homogeneous model, denoted as \mll(hom) throughout the paper (model UNREST in PAML documentation),  or the homogeneous time-reversible model denoted as \mll(homGTR). Rate matrix entries and root distribution had to be estimated by the software. We waited up to 60 seconds for convergence on each tree topology and if it did not converge, we treated it as failed (because we cannot compare likelihoods in this case). It is worth pointing out that, usually,  \ml was not convergent  only for the incorrect topologies.

In order to estimate HAL time-reversible model we used the software \bppml of the Bio++ package \citep{Dutheil2008} for the inference of HAL models with homogeneity across sites, \mll(HALGTR), and with discrete Gamma rates with two categories, \mll(HALGTR+2$\Gamma$).

As far as Neighbor-joining is concerned, the paralinear distance \citep{lake1994} was always used to estimate pairwise divergence.

\subsection{Simulations}
To generate data under the general Markov model, we have used
\texttt{GenNon-h} \citep{GenNonh}. Given a set of branch lengths and a tree topology, this software generates random root distribution and substitution matrices with the expected substitutions per site, and lets nucleotides evolve according to this Markov process on the tree.

In order to generate data evolving under homogeneous GTR model (with or without continuous Gamma-rates) we have used \texttt{Seq-gen} \citep{Rambaut1997}.  We used uniform root distribution, and the rate matrix underlying \texttt{Seq-gen} alignments on the tree space (Figure \ref{treespaceGTR} and Appendix 2.S2) had rates 2 (A$\leftrightarrow$C), 7 (A$\leftrightarrow$G), 4 (A$\leftrightarrow$T), 3 (C$\leftrightarrow$G), 1 (C$\leftrightarrow$T), 5 (G$\leftrightarrow$T), while the rate matrix underlying GTR+Gamma-rates had rates 2 (A$\leftrightarrow$C), 5 (A$\leftrightarrow$G), 3 (A$\leftrightarrow$T), 4 (C$\leftrightarrow$G), 1 (C$\leftrightarrow$T), 2 (G$\leftrightarrow$T).

\section{Acknowledgments}
We are indebted to B. Misof and C. Mayer for encouraging us to pursue this project and for their very useful comments. We wish to thank them and their group  for their warm  hospitality during our stay at the Alexander Koenig Zoological Museum.

Both authors are partially supported by
Spanish government MTM2012-38122-C03-01/FEDER and Generalitat de Catalunya 2009SGR1284.

\section{Author's contributions}
Both authors contributed equally to the development of this work.


\newpage

\section{Tables}
\FloatBarrier
{\linespread{1.1}
\begin{table}[h]
\begin{center}
\textbf{Average success of different quartet methods on the tree space of Figure 1b.}
\vspace*{5mm}

 \begin{tabular}{cc|cccc}
 \hline
  model & base pairs & \svd & \eri & \nj & \ml\footnotemark[1] \\ \hline \hline
   GMM & 1~000  & \textbf{0.856} (0.21) & 0.803 (0.17) & 0.797 (0.18) & 0.736 (0.17)\\
  & 10 000  & 0.958 (0.13)& \textbf{0.971} (0.04) & 0.943 (0.09) &  0.754 (0.17) \\ \hline
  homGTR & 500 & 0.732 (0.21) & 0.748 (0.22) & 0.729 (0.23) & \textbf{0.880} (0.11) \\
  & 1~000 & 0.796 (0.30) & 0.843 (0.19) & 0.805 (0.20) & \textbf{0.934} (0.06) \\
  & 10 000 & 0.940 (0.22) & \textbf{0.992} (0.04) & 0.945 (0.10) & 0.980 (0.02) \\ \hline
   \end{tabular}
   \end{center}
\caption{\label{tab:mean_sd} Average success (and standard deviation in parentheses) of \svd, \eri, Neighbor-Joining (\nj) and (homogeneous across lineages) maximum likelihood (\ml)  on data simulated on the tree space of Figure 1b according to the general Markov model (GMM) and the time-reversible model homogeneous across lineages and sites (homGTR) for different lengths and models (see Figure \ref{treespaceGMM}, \ref{treespaceGTR}, and Figure S2 in Appendix 2). In each row, the highest success is indicated in bold font.}
\end{table}
}
\footnotetext[1]{\mll(hom) is applied when data is generated under GMM (that is, it estimates the most general homogeneous continuous-time model), while \mll(homGTR) is applied when data is generated under the general time-reversible model homogeneous across lineages and sites.}

\FloatBarrier
{\linespread{1.1}
\begin{table}[h]
\begin{center}
\textbf{Quartet compatibility of \eri and \svd with the real data trees $T$, $T'$.}
\vspace*{5mm}

 \begin{tabular}{c|cccc}
 \hline
  topology & \svd & \eri ($m=1$) & \eri ($m=2$) & \eri ($m=3$) \\ \hline \hline
   $T$ & 91.43  & \textbf{84.29}  & \textbf{87.14} & \textbf{92.86} \\
  $T'$ & \textbf{94.26}  & 82. 86 & 77.14 & 85.71 \\ \hline
   \end{tabular}
   \end{center}
\caption{\label{tab:realdata} Percentage of quartets output by \svd and \eri (with different mixture assumptions) that are compatible with the yeast tree $T$ of \citet{Rokas2003} and the alternative tree $T'$ of \citet{Phillips2004}.  In each column, the highest success is indicated in bold font.}
\end{table}
}

\newpage

\section{Figures}
\pagestyle{empty}
\FloatBarrier


\newpage
\FloatBarrier
\appendix
\section{Supplementary Material}

\subsection{Appendix 1. Bipartition and transition matrices in the Felsenstein zone}

Here we illustrate the general situation on alignments corresponding to the Felsenstein zone. %

The two matrices on the following page correspond to the bipartition matrices $M_{12|34}(\tilde{p})$ and $M_{13|24}(\tilde{p})$ of an alignment generated under GMM on the tree of Figure 1.a with values $a=0.05$ and $b=1.01$ (Felsenstein zone).
%
In the second matrix, non-zero entries gather around the columns labeled with $AA$, $CC$, $GG$ and $TT$, while in the first matrix, we do not observe such an arrangement. This phenomenon is explained in terms of the short branch length between leaves $2$ and $4$, inducing few mutations between the sequences at these two leaves. This makes the matrix $M_{13|24}(\tilde{p})$ to be closer to rank-4 matrices than $M_{12|34}(\tilde{p})$. \eri corrects this bias by normalizing row and column sums so that all they have the same weight.

The following table displays the Frobenius distance of the bipartition matrices $M_{12|34}$, $M_{13|24}$ and $M_{14|23}$ to the space of matrices with rank $\leq 4$. According to these values, \svd would choose the topology $13|24$ as the right topology, as the corresponding value is the smallest among the values obtained by the 3 topologies. However, applying the correction introduced in  \eri, the topology $12|34$ is the one to be chosen as correct.

\begin{center}
\begin{eqnarray*}
\begin{tabular}{|c||c||ccc|} \hline
bipartition $A\mid B$ & $\mathrm{d}(M_{A\mid B})$ & $\mathrm{d}(M_{A\rightarrow B})$ & $\mathrm{d}(M_{B\rightarrow A})$ & $+$ \\
\hline \hline
$12|34$ & 0.000213291&  0.173455 & 0.336713 & \textbf{0.510168}
\\ \hline
$13|24$ & \textbf{0.0001063} & 0.479604 & 0.141812 & 0.621416
\\ \hline
$14|23$
& 0.000240727 &  0.17834 & 0.614431 & 0.792771
\\ \hline \hline
& \svd & & & \eri \\ \hline
 \end{tabular}
  \end{eqnarray*}
\end{center}

\begin{landscape}
\linespread{1}
\begin{scriptsize}
\begin{eqnarray*}
\begin{array}{c||cccccccccccccccc|c}
12|34 & AA & AC& AG & AT& CA & CC & CG & CT & GA & GC & GG & GT & TA & TC & TG & TT & + \\ \hline \hline
AA & 0.128 & 0.001 & 0.016 & 0.006 & 0.043 & 0.001 & 0.005 & 0.003 & 0.008  &  0 & 0.004  &  0 & 0.011  &  0  &  0 & 0.003 & 0.229 \\
AC & 0.001  &  0 & 0.001  &  0  &  0 & 0.002  &  0  &  0  &  0  &  0  &  0  &  0  &  0  &  0  &  0  &  0 & 0.004\\
AG & 0.001 & 0.001 & 0.021 & 0.001 & 0.002 & 0.002 & 0.027 & 0.001  &  0  &  0 & 0.009  &  0 & 0.001 & 0.001 & 0.001 & 0.002 & 0.070\\
AT & 0.002  &  0  &  0 & 0.011 & 0.001  &  0  &  0 & 0.013  &  0  &  0  &  0 & 0.002  &  0  &  0  &  0 & 0.039 & 0.068\\
CA & 0.023  &  0 & 0.002 & 0.003 & 0.005  &  0  &  0  &  0  &  0  &  0  &  0 & 0.001  &  0  &  0  &  0 & 0.001 & 0.035\\
CC & 0.001 & 0.025 & 0.004 & 0.002 & 0.003 & 0.140 & 0.011 & 0.003 & 0.001 & 0.026 & 0.008  &  0  &  0 & 0.012 & 0.001 & 0.006 & 0.243\\
CG & 0.001 & 0.003 & 0.028 & 0.002  &  0 & 0.0010 & 0.017  &  0 & 0.001 & 0.002 & 0.013  &  0 & 0.001  &  0 & 0.001 & 0.004 & 0.074 \\
CT & 0  &  0 & 0.001 & 0.019 & 0.001 & 0.002 & 0.001 & 0.029  &  0 & 0.001  &  0 & 0.006  &  0 & 0.003  &  0 & 0.055 & 0.118 \\
GA & 0.001  &  0  &  0  &  0 & 0.001  &  0  &  0  &  0 & 0.001  &  0  &  0  &  0  &  0  &  0  &  0  &  0 & 0.003 \\
GC & 0  &  0  &  0  &  0  &  0  &  0  &  0  &  0  &  0  &  0  &  0  &  0  &  0  &  0  &  0  &  0 & 0 \\
GG & 0  &  0 & 0.009  &  0  &  0 & 0.001 & 0.005  &  0  &  0  &  0 & 0.003  &  0  &  0  &  0 & 0.001 & 0.001 & 0.020\\
GT & 0  &  0  &  0 & 0.001 & 0.001  &  0  &  0 & 0.003  &  0  &  0  &  0  &  0  &  0  &  0  &  0 & 0.004 & 0.009 \\
TA & 0.010  &  0 & 0.001  &  0 & 0.001  &  0  &  0 & 0.001& 0.001  &  0 & 0.001 &  0  &  0  &  0 & 0.001  &  0 & 0.016\\
TC & 0  &  0  &  0 & 0.002  &  0 & 0.005  &  0  &  0 & 0.001  &  0  &  0  &  0  &  0  &  0  &  0 & 0.001 & 0.009\\
TG & 0 & 0.004 & 0.006  &  0  &  0 & 0.001 & 0.010  &  0  &  0  &  0 & 0.003 & 0.002  &  0  &  0 & 0.001 & 0.003 & 0.030 \\
TT & 0.002  &  0  &  0 & 0.010 & 0.001  &  0  &  0 & 0.020  &  0  &  0  &  0 & 0.001  &  0 & 0.001  &  0 & 0.036 & 0.071 \\ \hline
 + & 0.170 & 0.034 & 0.089 & 0.057 & 0.059 & 0.155 & 0.076 & 0.073 & 0.013 & 0.029 & 0.041 & 0.012 & 0.013 & 0.017 & 0.006 & 0.155 & 1
\end{array}
\end{eqnarray*}

\begin{eqnarray*}
\begin{array}{c||cccccccccccccccc|c}
13|24 & \textbf{AA} & AC& AG & AT& CA & \textbf{CC} & CG & CT & GA & GC & \textbf{GG} & GT & TA & TC & TG & \textbf{TT} & +\\ \hline \hline
AA & 0.128 &  0.001 &  0.016 &  0.006 &  0.001   &  0 &  0.001   &  0 &  0.001 &  0.001 &  0.021 &  0.001 &  0.002   &  0   &  0 &  0.011 &  0.190 \\
AC & 0.043 &  0.001 &  0.005 &  0.003   &  0 &  0.002   &  0   &  0 &  0.002 &  0.002 &  0.027 &  0.001 &  0.001   &  0   &  0 &  0.013 &  0.100\\
AG & 0.008   &  0 &  0.004   &  0   &  0   &  0   &  0   &  0   &  0   &  0 &  0.009   &  0   &  0   &  0   &  0 &  0.002  & 0.023\\
AT & 0.011   &  0   &  0 &  0.003   &  0   &  0   &  0   &  0 &  0.001 &  0.001 &  0.001 &  0.002   &  0   &  0   &  0 &  0.039  & 0.058 \\
CA & 0.023   &  0 &  0.002 &  0.003 &  0.001 &  0.025 &  0.004 &  0.002 &  0.001 &  0.003 &  0.028 &  0.002   &  0   &  0 &  0.001 &  0.019  & 0.114 \\
CC & 0.005   &  0   &  0   &  0 &  0.003 &  0.140 &  0.011 &  0.003   &  0 &  0.001 &  0.017   &  0 &  0.001 &  0.0020 &  0.001 &  0.029  & 0.213 \\
CG & 0   &  0   &  0 &  0.001 &  0.001 &  0.026 &  0.008   &  0 &  0.001 &  0.002 &  0.013   &  0   &  0 &  0.001   &  0 &  0.006  & 0.059 \\
CT & 0   &  0   &  0 &  0.001   &  0 &  0.012 &  0.001 &  0.006 &  0.001   &  0 &  0.001 &  0.004   &  0 &  0.003   &  0 &  0.055  & 0.084 \\
GA & 0.001   &  0   &  0   &  0   &  0   &  0   &  0   &  0   &  0   &  0 &  0.009   &  0   &  0   &  0   &  0 &  0.001  & 0.011 \\
GC & 0.001   &  0   &  0   &  0   &  0   &  0   &  0   &  0   &  0 &  0.001 &  0.005   &  0 &  0.001   &  0   &  0 &  0.003  & 0.011 \\
GG & 0.001   &  0   &  0   &  0   &  0   &  0   &  0   &  0   &  0   &  0 &  0.003   &  0   &  0   &  0   &  0   &  0  & 0.004 \\
GT & 0   &  0   &  0   &  0   &  0   &  0   &  0   &  0   &  0   &  0 &  0.001 &  0.001   &  0   &  0   &  0 &  0.004  & 0.006 \\
TA & 0.010   &  0 &  0.001   &  0   &  0   &  0   &  0 &  0.002   &  0 &  0.004 &  0.006   &  0 &  0.002   &  0   &  0 &  0.010  & 0.035 \\
TC & 0.001   &  0   &  0 &  0.001   &  0 &  0.005   &  0   &  0   &  0 &  0.001 &  0.010   &  0 &  0.001   &  0   &  0 &  0.020  & 0.039 \\
TG & 0.001   &  0 &  0.001   &  0 &  0.001   &  0   &  0   &  0   &  0   &  0 &  0.003 &  0.0020   &  0   &  0   &  0 &  0.001  & 0.009 \\
TT & 0   &  0 &  0.001   &  0   &  0   &  0   &  0 &  0.001   &  0   &  0 &  0.001 &  0.003   &  0 &  0.001   &  0 &  0.036 & 0.043 \\ \hline
+ & \textbf{0.233} & 0.002 & 0.030 & 0.018 & 0.007 & \textbf{0.210} & 0.025 & 0.014 & 0.007 & 0.016 & \textbf{0.155} & 0.016 & 0.008 & 0.007 & 0.002 & \textbf{0.249} & 1
\end{array}
\end{eqnarray*}
\end{scriptsize}
\end{landscape}

%

\newpage
\appendix
\subsection{Appendix 2. Supplementary material}


\vspace{5mm}

\subsubsection{\large{Performance of \svd}}

\begin{figure}[h]
\begin{center}
\includegraphics[scale=0.5]{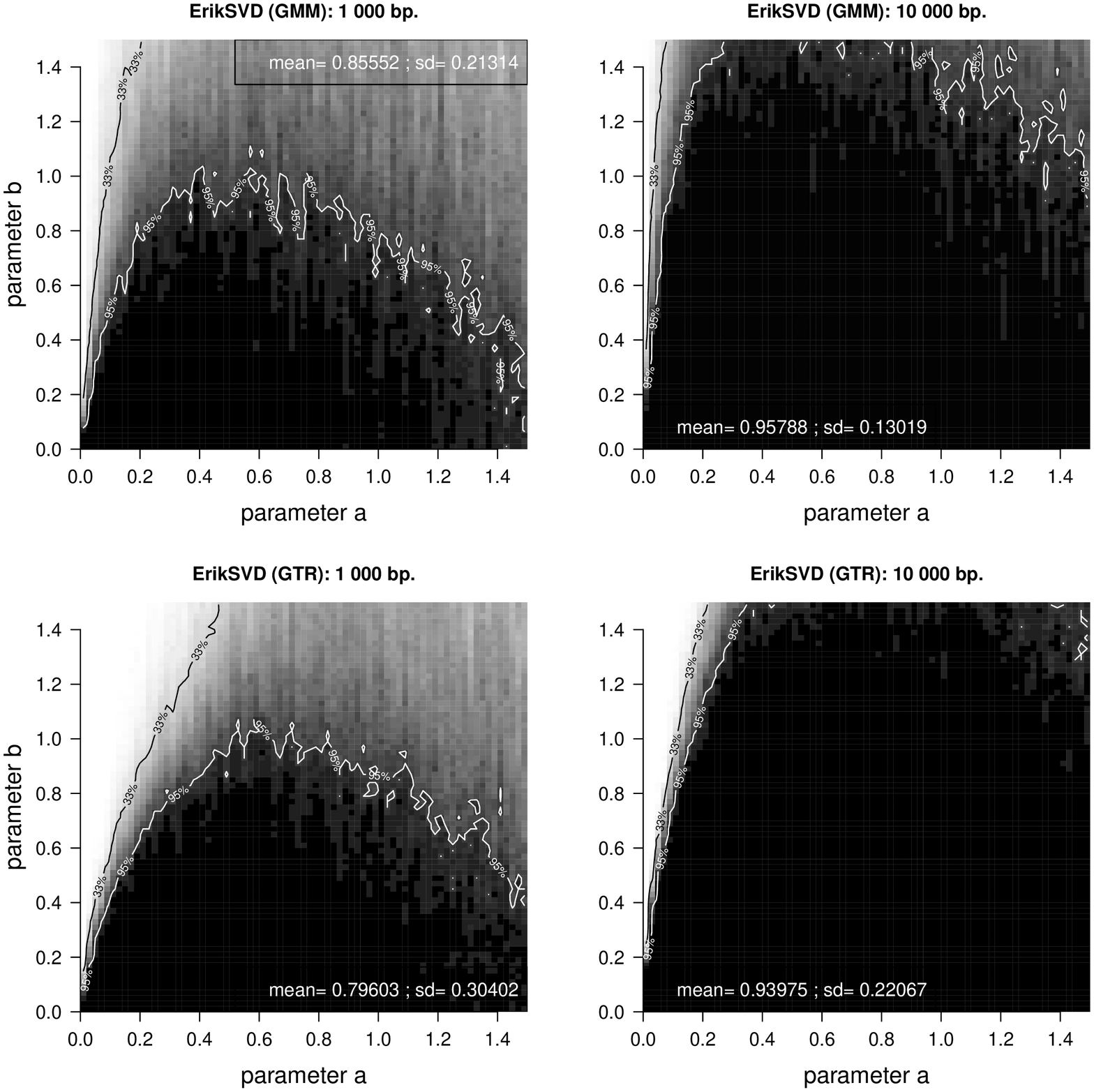}
\end{center}
\linespread{1.3}
Figure S1. Performance of \svd on the tree space of Figure 1.b  on data generated under GMM: black is used to  represent 100 \% of success, white to represents 0 \% and different tones of gray the intermediate frequencies.  %
Parameters $a$ and $b$ refer to the branch lengths of the tree of Figure 1.a, where $c$ is set equal to $a$.

\end{figure}

\newpage
\FloatBarrier

\subsubsection{\large{Performance on GTR data (500 bp.)}}

\vspace{5mm}
\begin{figure}[h]
\begin{center}
\includegraphics[scale=0.55]{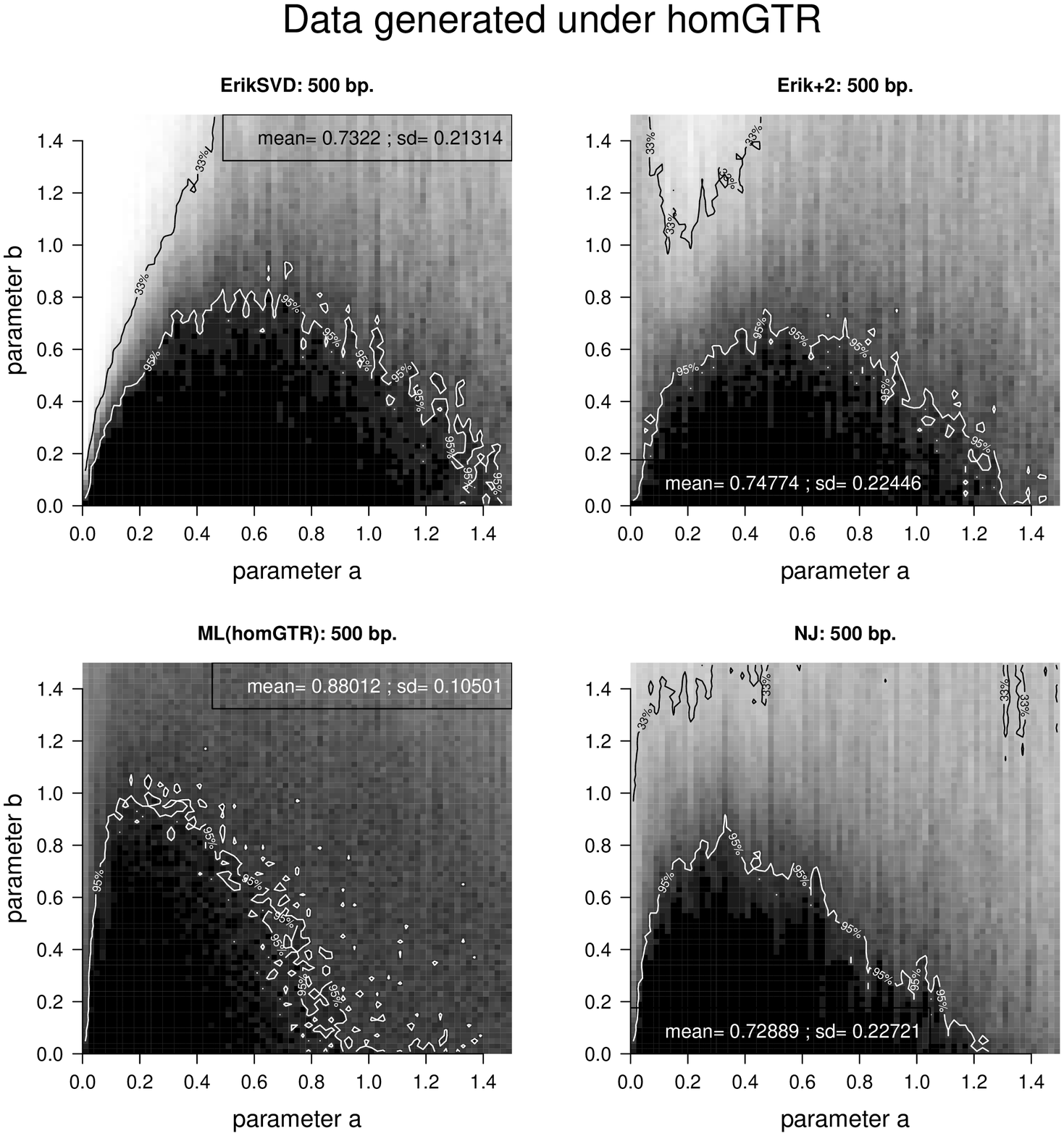}
\end{center}
\linespread{1.3}
\footnotesize
Figure S2. Performance  of \svd, \eri, \ml and \nj  on the tree space of Figure 1.b on alignments of 500 bp.  generated under homGTR; black is used to  represent 100 \% of success, white to represents 0 \% and different tones of gray the intermediate frequencies. 
%
%
\emph{Top Left}: \svd; 
\emph{Top Right}: \eri; 
\emph{Bottom Left}: ML estimating homogenous GTR model --ML(homGTR); 
\emph{Bottom Right}: Neighbor-Joining (paralinear distance). 
\end{figure}

\newpage
\FloatBarrier
\subsubsection{\large{Performance of \eri with $m=2$ on unmixed data}}

\vspace{5mm}
\begin{figure}[h]
\begin{center}
\includegraphics[scale=0.55]{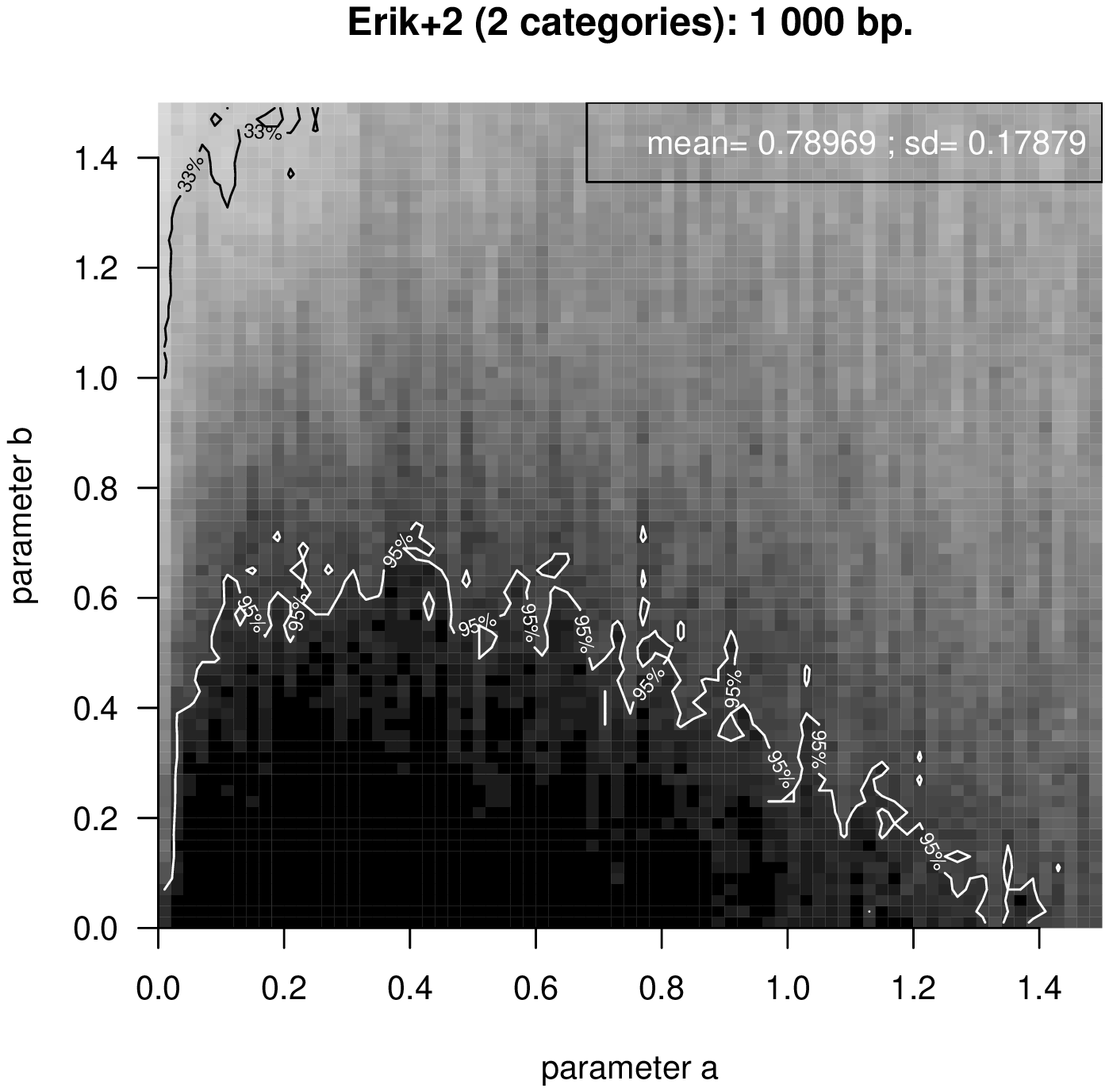}
\end{center}
\linespread{1.3}
Figure S3. Performance of \eri with $m=2$ on the tree space of Figure 1.b on data generated under GMM with homogeneity across sites for length 1 000; black is used to  represent 100 \% of success, white to represents 0 \% and different tones of gray the intermediate frequencies. %
Parameters $a$ and $b$ refer to the branch lengths of the tree of Figure 1.a, where $c$ is set equal to $a$. 
\end{figure}

\newpage

\subsubsection{\large{Performance of \eri with $m=1,2$}}

\vspace{5mm}
\begin{figure}[h]
\begin{center}
\includegraphics[scale=0.75]{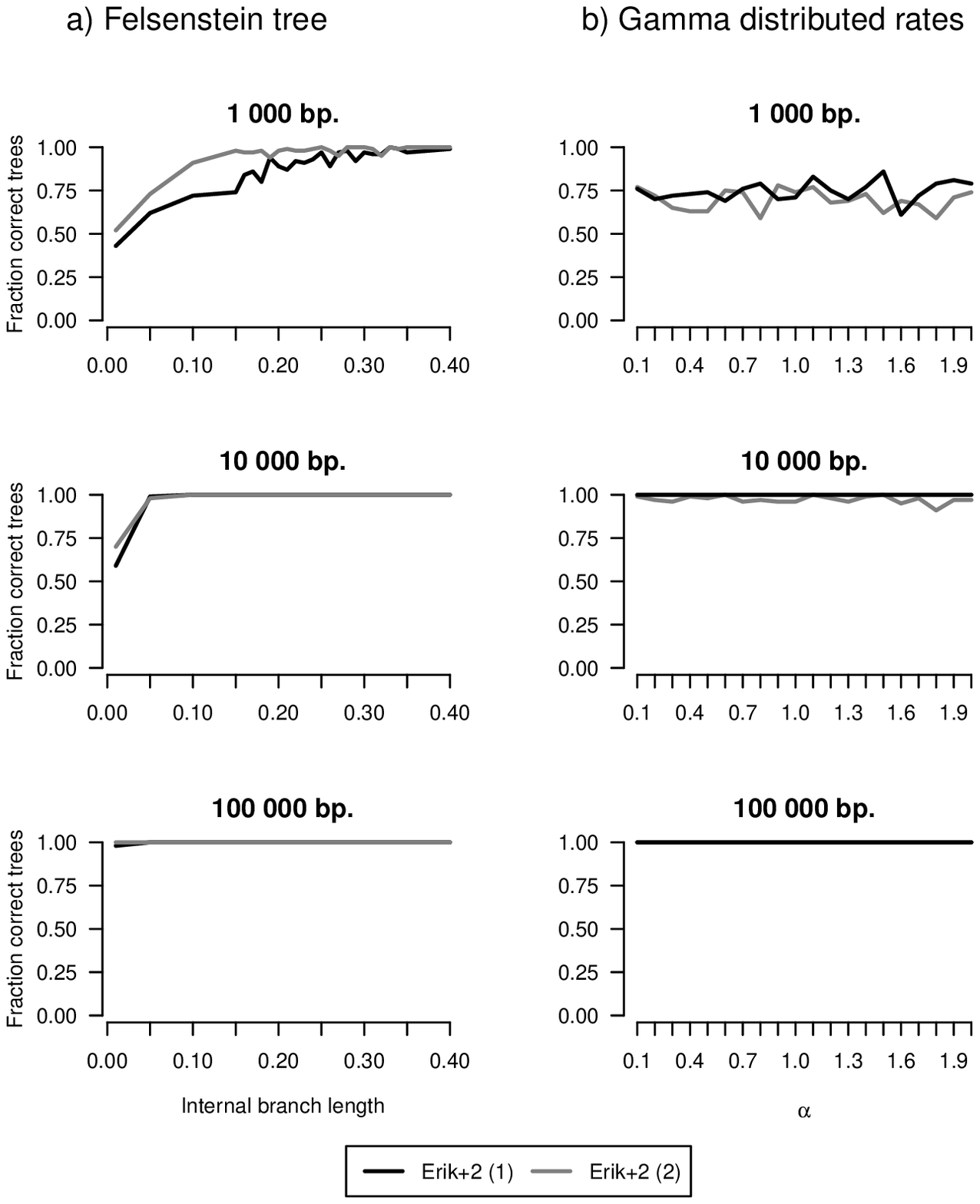}
\end{center}
\linespread{1.3}
\footnotesize
Figure S4. Percentage of correctly inferred trees by \eri with $m=1$ and $m=2$ on alignments of
lengths 1~000, 10~000 and 100~000 bp. shown from top to bottom. a)  Performance of \eri with $m=1$ and $m=2$ on data
generated under GMM on the tree of Figure 1.a with $a= 0.05$, $b=0.75$, and varying the internal
branch length $c$. 
b) Performance of \eri (with $m=1$) and \eri (with $m=2$) on data
generated under GTR model with continuous Gamma-rates and parameter $\alpha$ varying between 0.1 and 2 on the 4-taxa tree of Figure 1.a with branch lengths $a=c=0.05$ and $b=0.75$.
\end{figure}

\end{document}